\documentclass{sigcomm-alternate}
\usepackage{polski}
\usepackage[utf8]{inputenc}
\PassOptionsToPackage{hyphens}{url}\usepackage{hyperref}

\usepackage{enumitem}
\setlist{leftmargin=3.5mm,label={\raisebox{.45ex}{\rule{.6ex}{.6ex}}}}

\usepackage[font=bf]{caption}
\usepackage{subcaption}

\usepackage{etoolbox}
\newtoggle{thesis} 

\usepackage{color}
\usepackage[usenames,dvipsnames]{xcolor}

\hypersetup{
    colorlinks,%
    citecolor=BrickRed,
    filecolor=black,%
    linkcolor=blue,%
    urlcolor=BrickRed
}

\def\Snospace~{\S{}}
\usepackage{tikz}
\usetikzlibrary{arrows,positioning,shapes}
\usetikzlibrary{shapes,calc,shadows}
\pgfdeclarelayer{background}
\pgfdeclarelayer{foreground}
\newcommand\comment[1]{{\sffamily [xxx:  #1]}}
\newcommand\reviewfix[1]{{\sffamily [RF:#1]}}   
\newcommand\PostSubmission[1]{{\sffamily [post-xxx:  #1]}}
\iftrue
  \renewcommand\comment[1]{}
  \renewcommand\reviewfix[1]{}
  \renewcommand\PostSubmission[1]{}
\fi

\newlength{\onegraph}
\newlength{\twograph}
\newcommand{\LE}{\emph{LE~}}%
\newcommand{\LEs}{\emph{LE}}%
\newcommand{\LEf}{\emph{Let's Encrypt~}}%

\title{No domain left behind:\\is  Let's Encrypt democratizing encryption?}

\numberofauthors{3}
\author{
\alignauthor Maarten Aertsen\\
       \affaddr{Delft University of Technology\\The Netherlands}\\
       \affaddr{Maarten@rtsn.nl}
\alignauthor Maciej Korczy\'nski\\
       \affaddr{Delft University of Technology\\The Netherlands}\\
       \affaddr{ Maciej.Korczynski@tudelft.nl}
\alignauthor Giovane C. M. Moura\\
       \affaddr{SIDN Labs\\The Netherlands}\\
       \affaddr{Giovane.Moura@sidn.nl}
\and  
\alignauthor Samaneh Tajalizadehkhoob\\
       \affaddr{Delft University of Technology\\The Netherlands}\\
       \affaddr{S.T.Tajalizadehkhoob@tudelft.nl}
\alignauthor Jan van den Berg\\
       \affaddr{Delft University of Technology\\The Netherlands}\\
       \affaddr{J.vandenBerg@tudelft.nl}
}

\begin{document}
\maketitle
\begin{abstract}
	The 2013 National Security Agency revelations of pervasive monitoring have lead to an ``encryption rush'' across the computer and Internet industry. To push back against massive surveillance and protect users privacy, vendors, hosting and cloud providers have widely deployed encryption on their hardware, communication links, and applications. 
As a consequence, the most of web traffic nowadays is encrypted. However, there is still a significant part of Internet traffic that is not encrypted. It has been argued that both \textit{costs} and \textit{complexity} associated with obtaining and deploying X.509 certificates are major barriers for widespread encryption, since these certificates are required to established encrypted connections. To address these issues, the Electronic Frontier Foundation, Mozilla Foundation, and the University of Michigan have set up \textit{Let's Encrypt} (\LEs), a certificate authority that provides both free X.509 certificates and software that automates the deployment of these certificates. 
In this paper, we investigate \textit{if} \LE has been successful in democratizing encryption: 
we analyze certificate issuance in the first year of \LE and show from various perspectives that \LE adoption has an upward trend and it is in fact being successful in covering the lower-cost end of the hosting market.

%


\end{abstract}

\section{Introduction}
	The 2013 National Security Agency (NSA) revelations of pervasive monitoring and surveillance had a significant impact on the Internet industry. As a reaction, we have witnessed a surge on deployment of encryption technologies to curb these surveillance practices. For example, Google enabled encryption in the links between its datacenters~\cite{googledatacenter} while Apple enabled encryption by default on its mobile devices~\cite{appleEncryption}. The Internet Engineering Task Force (IETF) --a body that standardizes Internet-related protocols-- issued RFC~7258~\cite{rfc7258}, making it clear that ``pervasive monitoring is an attack''.

We have also seen a surge on the encryption of web traffic in response to these revelations. For example, browser telemetry from both Mozilla Firefox and Google Chrome shows that more than 50\%  of page loads by their users is currently encrypted~\cite{mozillatelemetry,transparencyreport}.
However, a significant portion of web traffic is still unencrypted, and it has been argued that both the complexity and costs associated with obtaining and deploying the required X.509 certificates (issued by third-party paid certificate authorities -- CAs) are major barriers for wide encryption of web traffic~\cite[p.86]{kasten2015server}. For example, some CAs charge 80~USD per certificate, per website per year and require manual setup.

To address these barriers against ubiquitous encryption, the Electronic Frontier Foundation (EFF), Mozilla Foundation, and University of Michigan set up \textit{Let's Encrypt}~\cite{letsencrypt-site} (\LE hereafter), a CA that provides both \textit{free} X.509 certificates and \textit{automated} software to configure servers to use those certificates.
 By reducing both costs (to zero) and deployment complexity, \LE aims to make encrypted traffic ubiquitous, democratizing certificate issuance and deployment.
Little after one year after launch,
\LE has issued 12 million certificates~--~making it to the top three largest CAs \cite{effblog, w3techblog}.

In this paper, we investigate \textit{if} \LE has been successful in democratizing encryption, 
and perform a comprehensive analysis on the issuance of \LE certificates. We use as a starting point one year of data obtained from the Certificate Transparency (CT) logs \cite{knownlogs}
and make the following contributions: looking from various perspectives, we show that \LE is indeed democratizing encryption -- we show that 98\% of certificates are issued for 
domains outside Alexa 1M (\autoref{sec:alexa}), but that issuance is not restricted to the lower-cost end of the market. Moreover, we show that the success of \LE is attributed by the adoption of major players (3 hosting providers are responsible for 47\% of the \LE certified domains, \autoref{sec:org-size}). We also show that issuance is dominantly for the lower-cost end of the market (shared hosting, \autoref{sec:shared}), and that the majority of certificates are correctly renewed after their first expiration (90 days, \autoref{sec:survival}). For the \url{.nl} top-level domain (TLD), we show that both old and new domains are benefiting from \LE (\autoref{sec:nl-cert-age}). Last, we show that 63\% of \LE certified domains are correctly deploying their certificates (\autoref{sec:issue-vs-deploy}), 
which is a lower bound number that we determined by performing active \texttt{https} scans. 
\vspace{-0.2cm}

\section{Background }
\label{sec:background}
	

\tikzstyle{int}=[draw, fill=blue!20, text width=3.5em, minimum width=5em, rounded corners, drop shadow]
\tikzstyle{tld} = [circle, fill=green!40, text width=2.5em, minimum width=3em, drop shadow, rounded corners]
\tikzstyle{db} = [cylinder,cylinder uses custom fill,cylinder body fill=red!20,cylinder end fill=red!20,      shape border rotate=90, aspect=0.25,draw,minimum size=3em]

\tikzstyle{server} = [draw, fill=orange!20, text width=4.5em, minimum width=3em, drop shadow, rounded corners]
\tikzstyle{outer}=[draw=gray,dashed,fill=none,thick,inner sep=5pt,rounded corners]
\tikzstyle{inner}=[draw=black,solid,fill=cyan!20,thick,inner sep=5pt,rounded corners]

\tikzstyle{site} = [circle, fill=orange!60, text width=2.5em, minimum width=3.2em, drop shadow, rounded corners]

\begin{figure}[t]
\centering

\tikzstyle{int}=[draw, fill=blue!20, text width=3.5em, minimum width=5em, 
rounded corners, drop shadow]
\tikzstyle{tld} = [circle, fill=green!40, text width=2.5em, minimum width=3em, 
drop shadow, rounded corners]
\tikzstyle{db} = [cylinder,cylinder uses custom fill,cylinder body 
fill=red!20,cylinder end fill=red!20,      shape border rotate=90, 
aspect=0.25,draw,minimum size=3em]

\tikzstyle{server} = [draw, fill=orange!20, text width=4.5em, minimum 
width=3em, drop shadow, rounded corners]
\tikzstyle{outer}=[draw=gray,dashed,fill=none,thick,inner sep=5pt,rounded 
corners]
\tikzstyle{inner}=[draw=black,solid,fill=cyan!20,thick,inner sep=5pt]

\tikzstyle{site} = [circle, fill=orange!60, text width=2.5em, minimum 
width=3.2em, drop shadow, rounded corners]

\begin{tikzpicture}[node distance=2.3cm,scale=0.65, every 
node/.style={scale=0.85}]

 \node[inner, minimum height=5em, minimum width=7em,] (ca) at (2.2,0) {\textbf{CA}} ;
 \node[inner, minimum height=5em, minimum width=7em,] (browser) at (11.2,0) {{\textbf{Browser/User}}} ;
 


\node[diamond,draw,align=center,text width=1.1cm,fill=yellow,drop shadow] 
(server) at (6.7,0) {\textbf{Web Server}};

 
  \draw[<-] ([yshift=-1 * 0.5 cm]server.west)  -- ([yshift=-1 * 0.5 cm]ca.east) node[midway,above] {\textbf{(ii)}};

 \draw[->] ([yshift=1 * 0.5 cm]server.west)  -- ([yshift=1 * 0.5 cm]ca.east) node[midway,above] {\textbf{(i)}};

  
 \draw[<-] ([yshift=0.8 cm]server.east)  -- ([yshift=0.8 cm]browser.west) node[midway,above] {\textbf{(iii)}};

 \draw[->] ([yshift=1 * 0 cm]server.east)  -- ([yshift=1 * 0 cm]browser.west) node[midway,above] {\textbf{(iv)}};
 
  \draw[<->] ([yshift=-0.8 cm]server.east)  -- ([yshift=-0.8 cm]browser.west) node[midway,above] {\textbf{(v)}};


\end{tikzpicture}%
	\caption{SSL/TLS connections and CAs}
	\label{fig:pkibasics}
	\vspace{-0.4cm}
\end{figure}
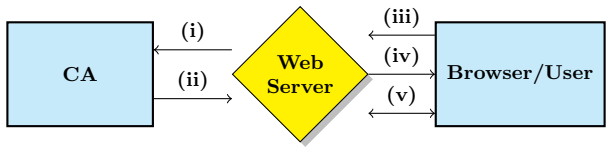

%
%

\subsection{SSL/TLS connections and CAs}
To illustrate how encrypted traffic on the Web works, consider the following example.
In \autoref{fig:pkibasics}, a user's browser connects to a web server to retrieve a webpage\footnote{X.509 certificates can actually be used for various  applications such as e-mail or \texttt{ssh}, but for the sake of simplicity, we only focus on web traffic here.}. 
After establishing the TCP connection, the client (browser) and server start the SSL/TLS handshake, which we briefly summarize here and refer the reader to \cite{rfc5246, infocom} for more details. The browser first sends a client hello message (iii), the server responds with a server hello message and a certificate message which includes its public key (iv). Upon receiving the certificate message, the browser must validate the chain of certificates \cite{rfc5280}, and only after this step the SSL/TLS setup continues and the encrypted connection can be used (v).

However, there are two prior steps necessary to get the required certificate: an entity has to request a  certificate from the CA for the particular fully qualified domain name (FQDN) (i). The CA, in turn, issues a certificate (ii), which is then deployed on the server. 

%


Commercial CAs typically offer three types of certificates: domain validated, organization validated, and extended validation certificates. All of them employ the same encryption measures -- they differ on how the CA verifies the 
user's identity (e.g.\ if the 
user is the legal owner of the domain and company 
for which a certificate is being issued).

Since \LE automates issuance, it only provides domain validated certificates, where a user merely has to prove ownership of the FQDN being certified. \LE is the first CA to fully automate 
the process of validation and issuance, using the Automatic Certificate Management Environment (ACME) protocol \cite{draft-ietf-acme}.

\subsection{Web hosting}
\label{sec:webhosting}

Web hosting is the industry that maintains content on the Web on behalf of customers (either paid or free). Web hosting is offered in  various types of  services such as shared versus dedicated hosting, according to customer's need and provider's business plan. In this paper, we are more interested in studying the adoption of \LE the in lower-cost end of the market, which employs shared hosting.

In shared hosting, a large number of websites are hosted on the same physical server, typically associated with less popular websites, allowing them to share the costs. To identify shared hosts, we use the methodology some of us introduced in~\cite{tajalizadehkhoobapples}, in which an IP address is classified as shared hosting if it hosts more than 10 unique domain names.%

\section{Datasets}
\label{sec:method-data}


\subsection{Certificate Transparency logs}
\label{sec:transparencyreports}
The certificates issued by \textit{LE} were obtained from Certificate Transparency (CT) logs~\cite{knownlogs}, which provide an append-only log of certificate issuance \cite{rfc6962-bis-19}.
\LE issued its first certificate on Sept 2015. Our data therefore contains
one year of certificates based upon CT data (Sept 2015-2016). Note that we only consider non-expired certificates given that an \LE certificate expires after 90 days and requires further renewal.

For each certificate, we extract their respective FQDNs from the \texttt{subjectAltName} string. We then transform these FQDNs into a ``normalized" \textit{domain} form, which is defined as either the  $2^{nd}$--level or $3^{rd}$--level 
if a given TLD registry provides such registrations (e.g.: \url{example.co.uk} or \url{example.org}). Therefore, we \textit{do not} analyze the number of certificates issued by \LE in this paper, we focus on their coverage of their ``normalized" \textit{domain} form. For instance, certificates for \url{a.example.org} and \url{b.example.org}  would be 
mapped into one \textit{domain} (\url{example.org}). In the rest of this paper, we use domains in the sense of ``normalized" domains.

%



\subsection{DNSDB}
\label{sec:dnsdb}

The CT logs  only provide information about the domains that have been issued certificates; it does not include information about  \textit{where} these domains are hosted.  
To determine that, we need to rely upon DNS data to check which IP addresses are associated with each of the domains in our~data.


Specifically, we use passive DNS logs (Sept 2015-2016) obtained from DNSDB -- a passive DNS database provided by Farsight Security \cite{dnsdb}.
To our knowledge, DNSDB has the best coverage of the overall domain name space that is available to researchers.
We found that DNSDB contains historical A records\footnote {A records are type of DNS records that map domains into IP addresses~\cite{rfc1034}.} for 80\% of \LE domains. An alternative to DNSDB would be to have performed DNS lookups for all domains covered by \LE\hspace{-0.12cm}, 
but the required historical data for that purpose was not available.

Therefore, we employ DNSDB as a historical \textit{sample} of the complete domain space, covering roughly 127--205 million unique domains monthly. 

\subsection{Organizations mapping and classification} 
\label{sec:orgmapping}

We map the IP addresses obtained from DNSDB into the corresponding organizations with which they are hosted.  In short, for each IP address, we retrieve the organization that this IP is assigned to (or  is ``allocated'' to), using their respective RIR IP \texttt{whois} data and historical Maxmind GeoIP2 database~\cite{maxmind-org}. We then use the methodology and keywords described in our previous work~\cite{tajalizadehkhoobapples} to classify the organizations according to their business plans: hosting providers, content delivery networks (CDNs), educational, among others.



\subsection{.nl domain registration information}
\label{sec:sidn}

To determine the domain age of certified \LE domains, we employ registration information provided by the \url{ .nl} TLD registry (SIDN). An alternative was to analyze \texttt{whois} records for all the other zones. However, given the fact that (i) most of TLDs do not offer historical domain \texttt{whois} service, (ii) it cannot be publicly accessed,
(iii) and for those that do offer \texttt{whois}, the format is not standardized~\cite{Liu:2015:CLP:2815675.2815693}, we opt for singling out \url{.nl} as a case study.

\section{Analysis and Discussion}
	\label{sec:results}
	
\subsection{Absolute and relative growth}
\label{sec:domains}
How big is \LEs?
\LE publishes statistics \cite{letsencryptstats} showing a  continuous growth in the number of daily issued certificates. \LE is in fact the third biggest CA, according to other research~\cite{w3techblog} and most of its growth comes from sites that did not have certificates before~\cite{effblog}.

\begin{figure}
\centering
 \includegraphics[width=\onegraph]{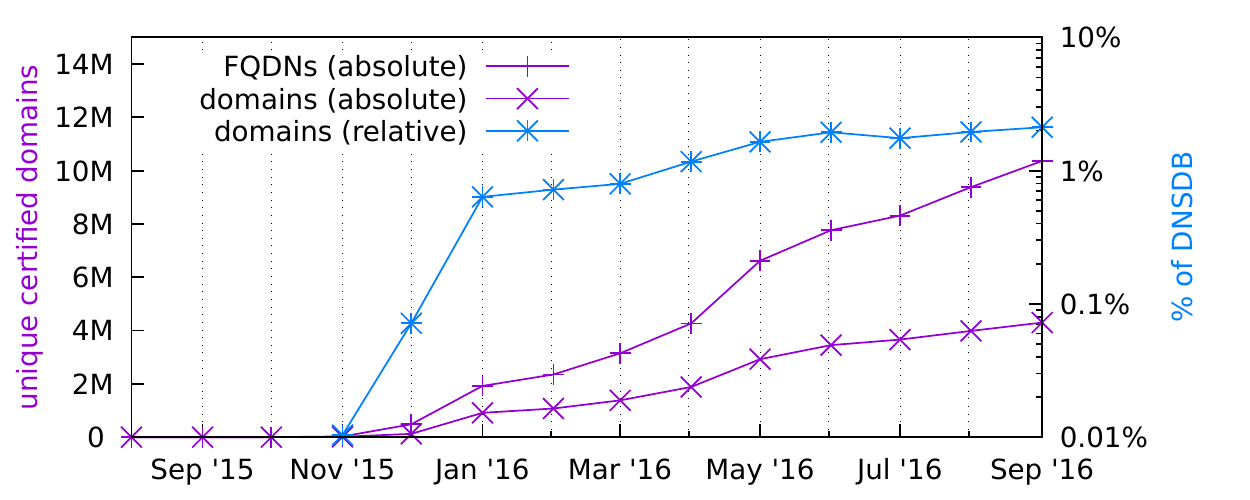}
\caption{\LE time series for FQDNs, domains, and DNSDB ratio}
\label{fig:descr-domains}
\vspace{-0.3cm}
\end{figure}

\autoref{fig:descr-domains} shows a time series of the absolute number of unique \LE certified domains, FQDNs, and domains relative to all domains observed in DNSDB~(\autoref{sec:dnsdb}). First, we see a continuous growth in all metrics: by Sept 2016, there were $\sim$10.4M FQDNs that had \LE certificates, amount to  $\sim$4.3M
domains (\autoref{sec:transparencyreports}), on average 2.5 certificates per domain.

Moreover, to have an idea on how much of the domain name space uses \LE certificates, we use DNSDB as a comparison and show that  \LE is used by 2\% of all domains observed in Sept 2016. Given that \LE has been only active for a year by the time of this analysis, 2\% of a large sample of the domain space represents a significant growth.


\begin{figure}[ht!]
\centering
 \includegraphics[width=\onegraph]{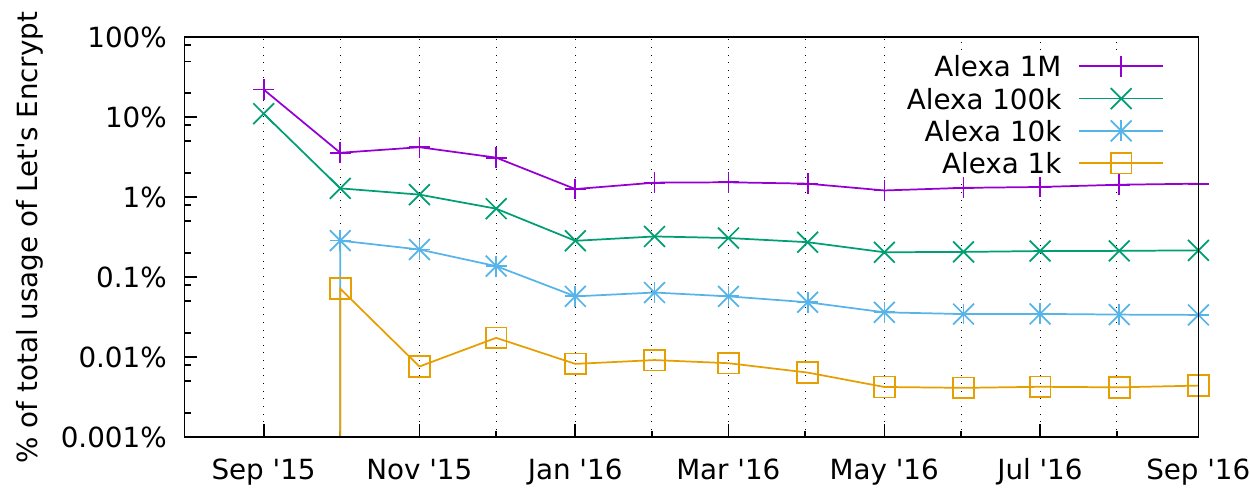}
        \caption{\LE certificates issued to Alexa top-ranked domains}

\label{fig:descr-alexa}
\vspace{-0.3cm}
\end{figure}

\subsection{Popular sites and \large \textbf{\LEf}}
\label{sec:alexa}

Intuitively, one could think that \LE would be predominantly used among less popular domains and that popular domains would already have their paid certificates deployed, therefore not using free \LE certificates. In this subsection, we examine this assumption.
To do that, we obtain the list of most popular websites from Alexa \cite{alexa} and match \LE domains against it.

\autoref{fig:descr-alexa} shows a time series of the relative contribution of Alexa ranked domains (1M, 100K, 10K, 1K) against the total number of domains with valid \LE certificates. The contribution of Alexa 1M domains remains stable around $2\%$ of total \LE usage throughout 2016, a period of relatively rapid growth of  \LE issuance.
By Sept 2016 about~$\sim$64K domains ranked in the Alexa 1M use \LE (\autoref{sec:transparencyreports}).

\begin{figure}[ht!]
\centering
 \includegraphics[width=\onegraph]{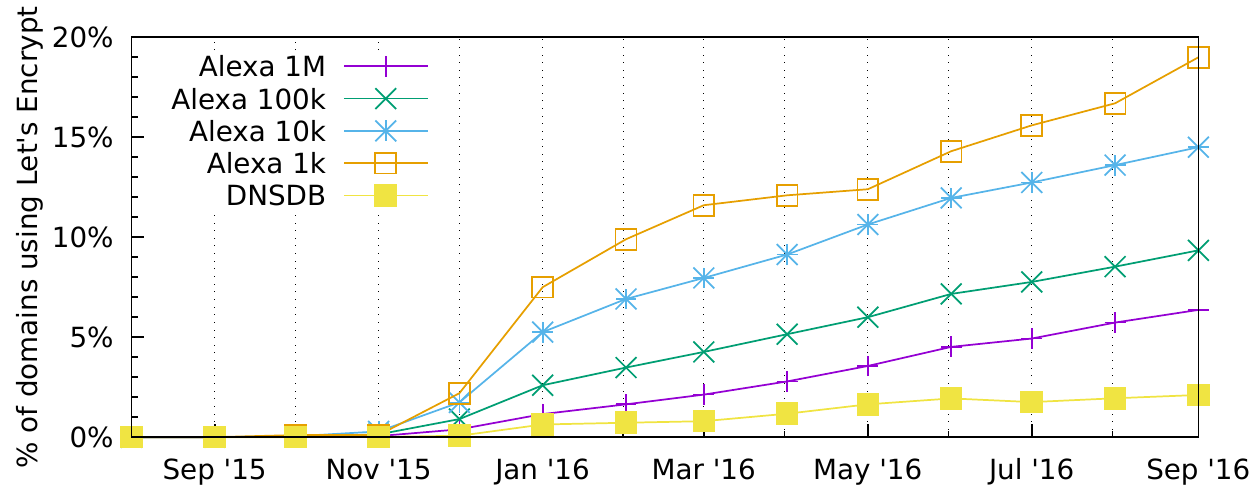}
	\caption{Relative usage of \LE domains in Alexa rankings}
\label{fig:descr-alexa2}
\vspace{-0.3cm}
\end{figure}

\autoref{fig:descr-alexa2} shows a time series of the relative growth of issuance within the Alexa rankings. By Sept 2016, 19\% of domains in the Alexa 1K had at least one certificate issued for a FQDN under their domain (e.g. \url{subdomain.example.org}).
This suggests that admins of 19\% of the most popular websites know about \LEs's existence and use its service, but they do not necessarily issue and deploy certificates on their main websites (e.g., both \url{wsj.com} and \url{welt.de} are labeled as \LE domains, yet do not use \LE on their main websites).

Overall, we find that 98\% of the certificates are issued for less popular sites outside Alexa rankings -- which is good for democratizing encryption -- and 2\% are issued for popular and resourceful websites, indicating that \LE is not only constrained to the lower-cost share of the market.

\subsection{Certificates distribution per organization}
\label{sec:org-size}

Which organizations are using more \LE certified domains? Are there ``big players'' or the \LE domains are distributed across small organizations? 
To answer this question, we use the methodology described in \autoref{sec:orgmapping} and map \LE domains to their respective IP address owners. 
We calculate~size indicators, by aggregating these mappings per organization.

\begin{figure*}
\centering
	\begin{subfigure}[b]{0.25\twograph}
		\includegraphics[width=\textwidth]{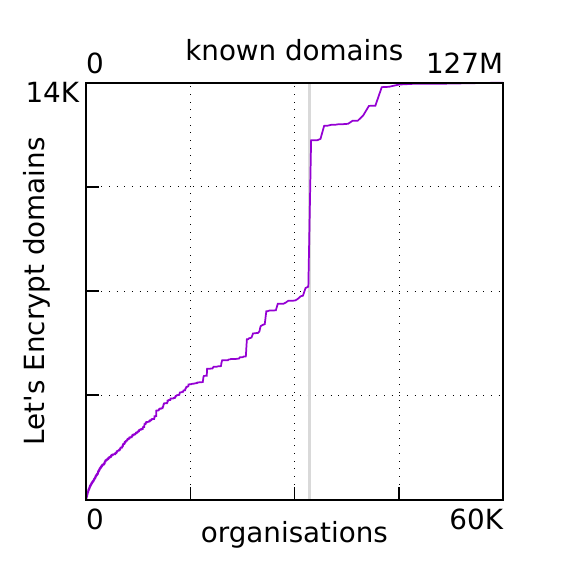}
		\caption{November 2015}
	\end{subfigure}
\hfill
	\begin{subfigure}[b]{0.25\twograph}
		\includegraphics[width=\textwidth]{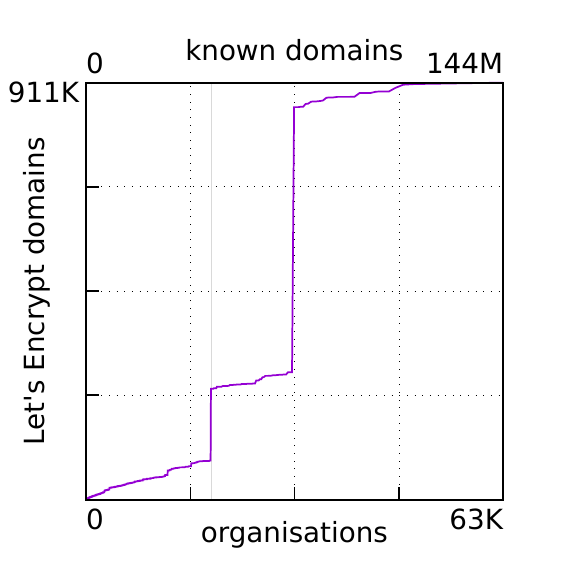}
		\caption{January 2016}
	\end{subfigure}
\hfill
	\begin{subfigure}[b]{0.25\twograph}
		\includegraphics[width=\textwidth]{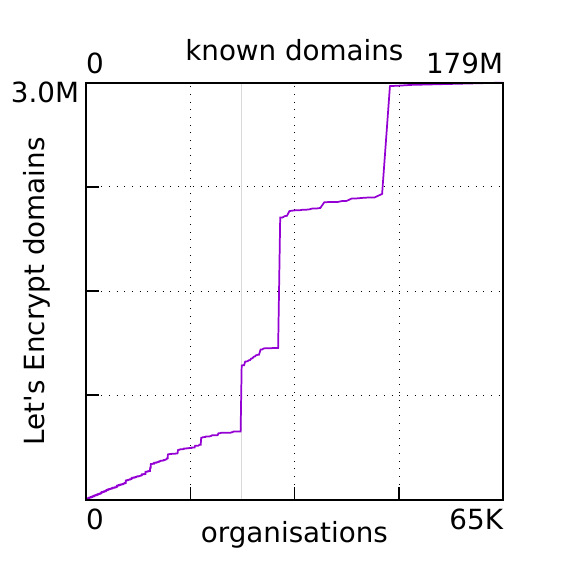}
		\caption{May 2016}
	\end{subfigure}
\hfill
	\begin{subfigure}[b]{0.25\twograph}
		\includegraphics[width=\textwidth]{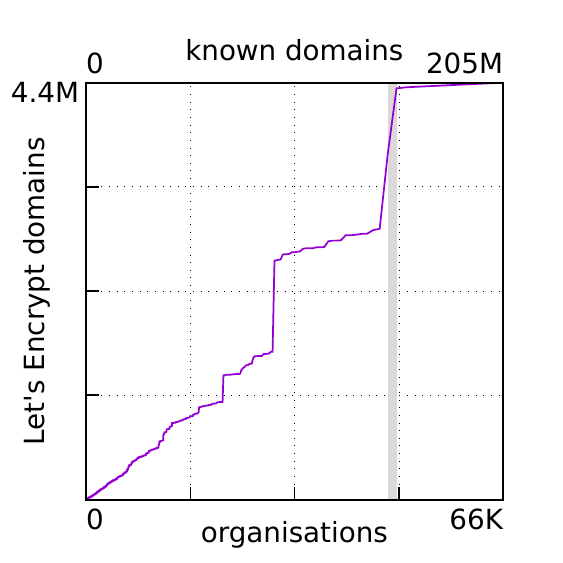}
		\caption{September 2016}
		\label{fig:descr-org-size-nov}
	\end{subfigure}
\caption{ECDF of \LE use versus organization size (domain density, measured in number of associated domains).
The x-axis on bottom has organizations sorted by domain density in ascending order. The y-axis represents the total number of \LE domains issued that month. The x-axis on top represents the total domains in DNSDB
The shaded area indicates domains that are not successfully attributed to an organization.}
\label{fig:descr-org-size}
\vspace{-0.1cm}
\end{figure*}

\autoref{fig:descr-org-size} shows ECDF of \LE certified domains per organization, for four selected months of issuance. We sort the organizations ($x$ axis) by their size in terms of domains hosted on allocated IP addresses. Steps in these lines indicate bulk issuance of \LE domains by an organization.
For example, in Jan 2016, we see the large vertical line corresponding to deployment at Automattic/wordpress.com ($x=0.5, \Delta y=63.5\%$), which is especially noticeable when compared against Nov 2016.
By Sept 2016, we can observe three clear steps: Shopify ($x=0.33, \Delta y=6\%$), Automattic/wordpress.com ($x=0.45, \Delta y=22\%$) and OVH ($x=0.7, \Delta y=19\%$). All three companies have announced issuance for their customers and are jointly responsible for 47\% of \LE certified domains. It is exactly these companies, serving numerous, smaller customers that would otherwise not enable the use of encryption by their visitors.

We also find evidence which suggests that \LE benefits smaller organizations.
Among all 66K identified organizations using the methodology explained in \autoref{sec:orgmapping}, we find 14K that have domains certified with \LE in Sept 2016.
Notably, 9K have 5 or less \LE certified domains. This corresponds to the lower left quadrant of~\autoref{fig:descr-org-size-nov}, where smaller organizations are jointly responsible for 
23\% of all \LE domains.

We conclude that \LE reaches both large hosting companies as well as smaller organizations with lower domain concentration.

\subsection{Types of organizations}
\label{sec:org-type}

In the previous section, we analyzed the distribution of \LE certified domains per organization. In this section, we classify these organizations according to their types 
(\autoref{sec:orgmapping}, \cite{tajalizadehkhoobapples}).
We group organizations into government related, educational, domain parking, hosting providers, ISPs, CDNs, DDoS-protection services and others.

\begin{figure}
\centering
 \includegraphics[width=\onegraph]{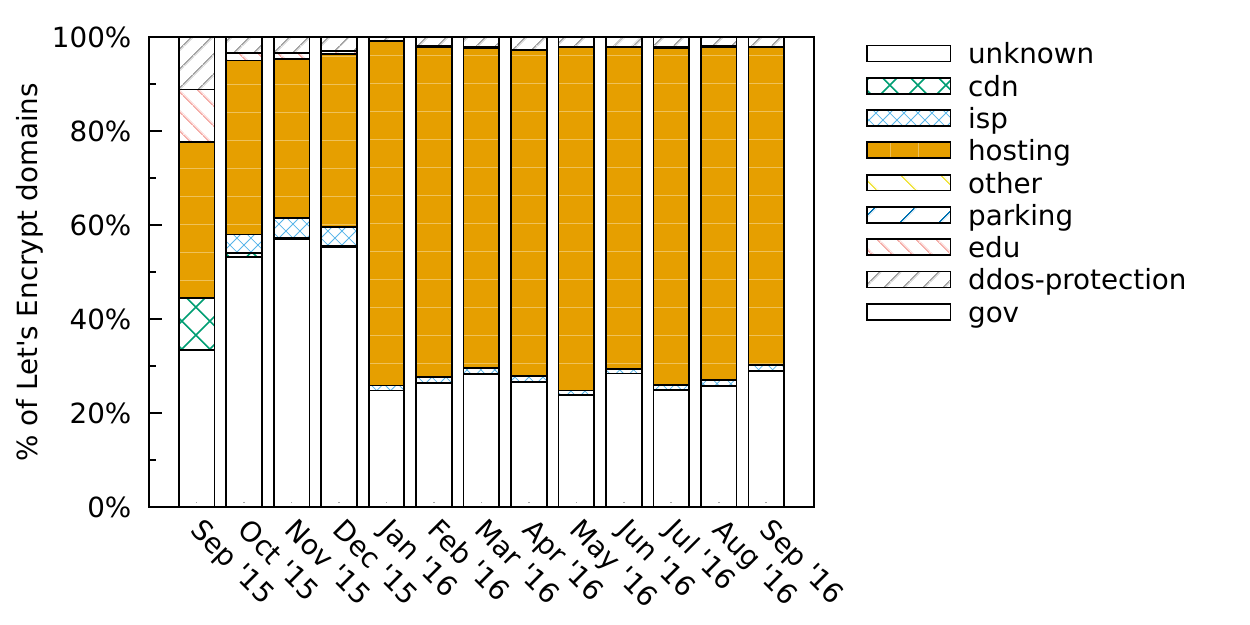}
\caption{Distribution of \LE domains per organization type (in \% of domains)}
\label{fig:descr-org-type}
\vspace{-0.2cm}
\end{figure}

The distribution of \LE domains per organization category is shown in \autoref{fig:descr-org-type}. 
As expected, the majority of domains are associated with hosting organizations (68\% in Sept 2016), while the share of DDoS protection services and CDNs remains low (2\% and 0.1\%, respectively). 
Note, however, that 29\% of all domains were not attributed to any of the categories (`unknown').

\subsection{Types of hosting: shared and non-shared}
\label{sec:shared}

Hosting services typically are offered in multiple types at different prices. 
Resources such as CPU, memory, bandwidth and IP addresses could be dedicated to customers (``dedicated hosting''), or shared among them. Shared hosting is where prices are at their lowest level and profit margins are slimmer. Under these conditions, encryption deployment would be least expected.


We classify the IP addresses from the hosting organizations listed in \autoref{sec:org-type} into shared and non-shared hosting via the methodology explained earlier in \autoref{sec:webhosting} \cite{tajalizadehkhoobapples}. 
%

\begin{figure}
\centering
 \includegraphics[width=\onegraph]{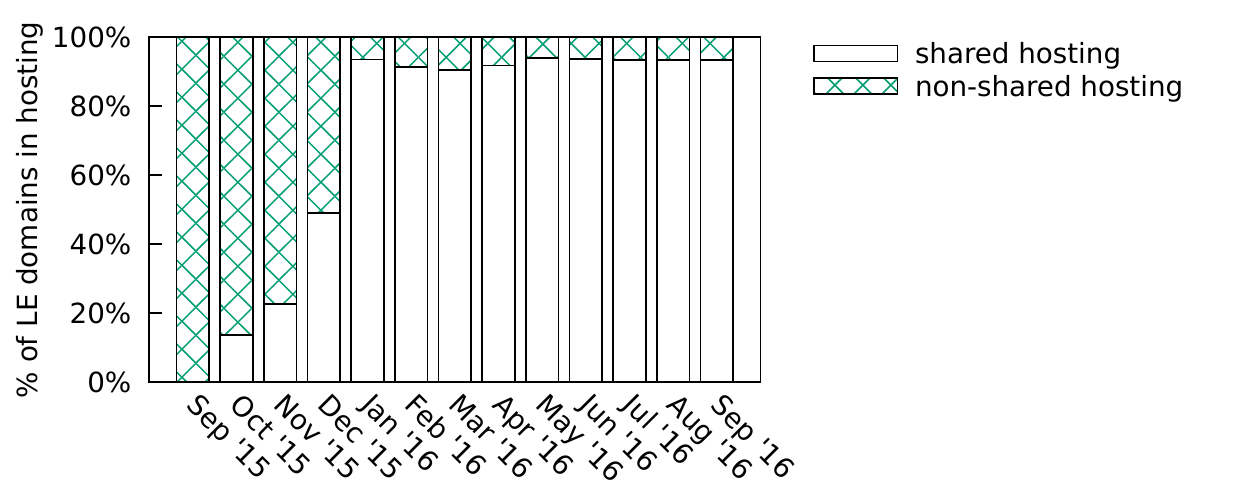}
	\caption{Distribution of \LE accross shared and non-shared hosting (in \% of domains)}
\label{fig:descr-shared}
\vspace{-0.45cm}
\end{figure}

\autoref{fig:descr-shared} is a histogram of relative market share within the hosting segment, split between shared and non-shared hosting services. We find that from Jan 2016, \LE use within hosting is dominantly connected to shared hosting services, with a penetration above 90\%.
Recalling that by Sept 2016 the overall hosting segment is dominant over other types (67\%), we find that \LE has very high overall utilization in shared hosting, which has traditionally been the least likely candidate for adoption of encryption due to the associated costs. As in the previous section, we can see that \LE covers the lower-cost end of the market.
\subsection{Certification lifetime}
\label{sec:survival}

Once domains issue an \LE certificate, do they keep on renewing them every 90 days? Or do they let them expire? To answer this question, we carry out a survival analysis of \LE certificates for each FQDN using  a Kaplan-Meier Survival Estimate \cite{kaplan1958nonparametric}. We identify three components that are likely to influence the outcome of this question:
%
(i) renewal automation working correctly 
(not having automation set-up likely causes renewal failure);
(ii) user satisfaction with the service and its certificates;
(iii) the intended lifetime of the domains themselves.
%



\begin{figure}
\centering
 \includegraphics[width=\onegraph]{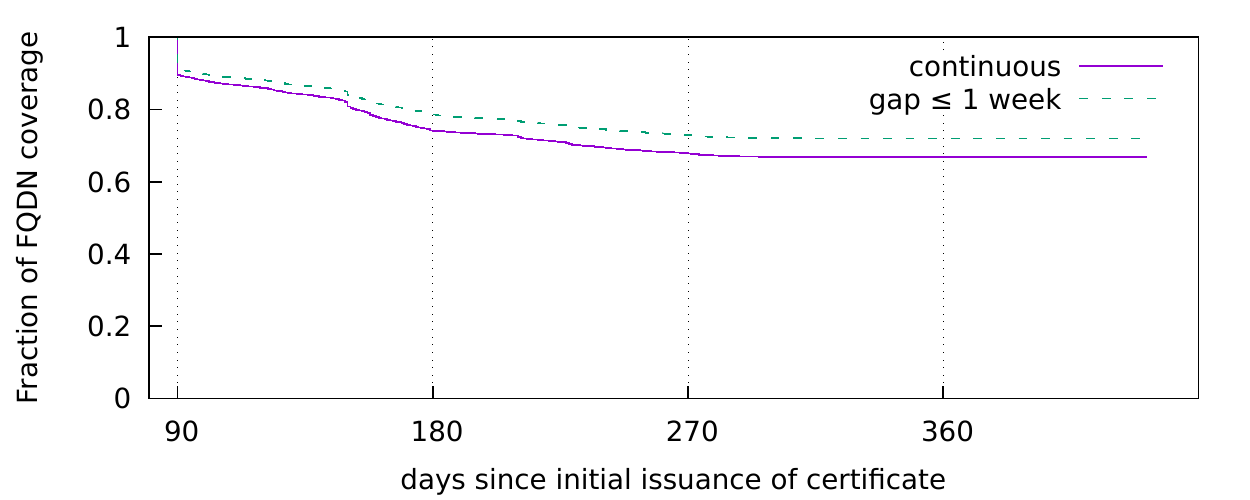}
\caption{Survival analysis of \LE certificates}
\label{fig:survival}
\vspace{-0.2cm}
\end{figure}

\autoref{fig:survival} shows the estimated survival function of \LE certified FQDNs featuring two functions. The continuous function measures survival without any downtime: survival implies the issuance of certificates with perfectly overlapping validity periods. The second function measures survival with a maximum one week gap in between consecutive validity periods. This accounts for 
failure in automation, corrected after the previous certificate expires. 

Since all certificates are valid for 90 days, we  observe 100\%, survivability for this period. After those 90 days we see drops: domains that either stop being re-certified, where automation was not successful or that the domain itself expired. The survival curve noticeably flattens after $x=270$~days, indicating that the automation is effective.

The agreement between $gap=0$ (continuous) and $gap \leq1~week$ indicates that beyond initial downtime, further survival is roughly similar. This may be explained by users that get continuous coverage after successful setup of automation. With more than $70\%$
FQDN survival after a full year, we can conclude that the 
majority of \LE users remain loyal to the service during our measurement
 period, which is not surprising given the size (\autoref{sec:org-size}) and type (\autoref{sec:org-type}) of \LE users -- dominantly (big) hosting providers.


\subsection{New vs. old domains}
\label{sec:nl-cert-age}

What type of domains are more likely
to employ \LE certificates: newly registered domains or older domains?
Since \LE reduces the certificate cost to zero while provides full
automation, one could hypothesize that registrars simply enable them by
default on their registrations system, so every new domain could be
automatically configured with an \LE certificate.

To investigate this hypothesis, we focus on the \url{.nl} TLD as a case study (\autoref{sec:sidn}).
There were 514,986 \LE certificates issued for 191,176 unique FQDNs, during the monitoring period.
In total, there were 85,223 unique \url{.nl} domains that had \LE certificates.

\begin{figure}
  \centering
   \includegraphics[width=\onegraph]{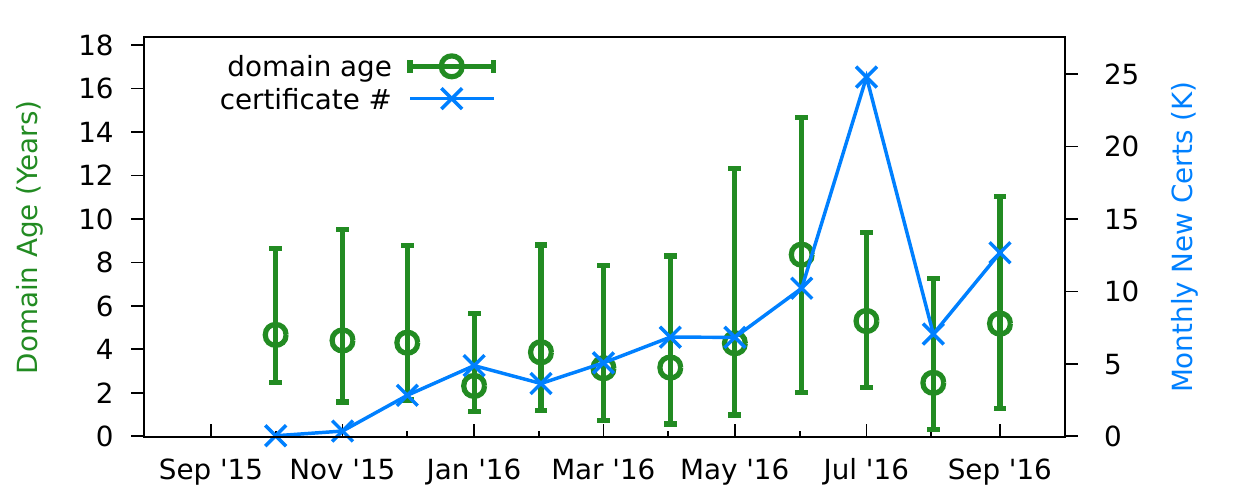}
   \caption{Median, Q25, Q75 and number of monthly new certificates for \url{.nl} domains}
   \label{fig:nl-cert-age}
   \vspace{-0.2cm}
\end{figure}

\autoref{fig:nl-cert-age} shows the number of \LE certificates issued for \url{.nl} domains for the first time (continuous line), and the median age, first quartile (Q25) and third quartile (Q75) (box plot). 
As we can see, for all months, the median age
of the domains is above two years, with a large spread, suggesting that \LE is
being used both for older and newer domains.  We can conclude that for
this dataset, most of \LE certificates are being used on already existing domains. In the absence of scan data for those domains, we cannot confirm if they had their first certificate
issued by \LEs, or if they switched to \LEs. However, it has been discussed in~\cite{effblog} that most \LE domains did not support encryption prior to their \LE certificate issuance.

%

\subsection{Certificate issuing vs. deployment}
\label{sec:issue-vs-deploy}

So far we have covered the side of certificate issuance. However, another open question is to determine how many of these certificates are actually deployed. Answering this question is not straightforward: first, certificates can be used for other applications than the Web, such as e-mail or \texttt{ftp}. Certificates can also be deployed internally within networks or be used on non-standard ports.

To have a lower-bound estimate of \LE certificate deployment, we randomly select 25,000 FQDNs for certificates that were valid between Nov 13 and 19, 2016 and scan them on \texttt{https} (TCP port 443) to determine if the certificates are actively deployed for use on the Web. We perform the scans on Nov 28, 2016.

\begin{figure}
  \centering
   \includegraphics[width=\onegraph]{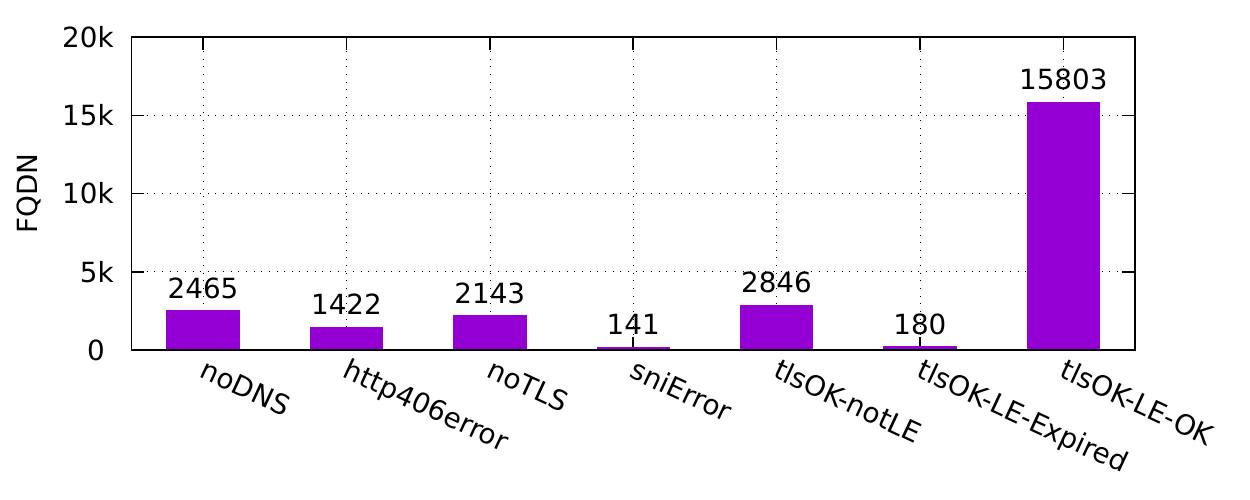}
	\caption{Scans of 25,000 \LE covered FQDNs}
   \label{fig:scans}
   \vspace{-0.2cm}
\end{figure}

\autoref{fig:scans} shows the scan results. As can be seen, 15,803 (63\%) of FQDNs have successfully deployed \LE certificates. The remaining were divided into other errors, such as 2,465 (10\%) having no DNS records -- e.g. short-lived, possibly expired; 2,143 (9\%) do not support TLS and 1,422 (6\%) return an \texttt{http} error code and are likely not set-up for \texttt{https} in the first place. Interestingly, 2,846 (11\%) deploy certificates not issued by \LEs. Here one could hypothesize that either the hosting provider is waiting for paid certificates to expire or is just experimenting. In addition, 180 FQDNs had expired \LE certificates.

Our results show that 63\% of our sampled FQDNs (as a lower-bound value) have successfully deployed \LE certificates. For a more comprehensive view on \LE deployment, it is important to 
perform longitudinal active measurements over all FQDNs covered by \LEs.

%


\section{Related work}
\label{sec:related}
	
The ecosystem for certificates and their use has been analyzed by various studies, but none of them have singled out \LE and analyzed its impact. For example, there are Internet-wide scan studies covering certificates~\cite{holz2011ssl,levillain2012one,durumeric2013analysis,censys15}. Several methods with the goal of mapping the CA ecosystem (e.g.: active scans, Certificate Transparency~\cite{knownlogs}) have also been compared~\cite{vandersloot2016towards}. Paid access reports have been previously issued (e.g. \cite{w3techdaily}). 



Although \LE is a new player in the CA market, there have been some preliminary efforts in measuring 
its adoption. 
However, none of them have been peer-reviewed at the time of the writing this paper. 
For example, there are self-reported \LE statistics pages (e.g. \cite{letsencryptstats}), a blog post by J.C. Jones~\cite{jones2016le}, and another one by Helme on the \LE coverage on the Alexa 1M~\cite{helmeAlexa}. The EFF compared the size of \LE against other CAs in 
their blog post~\cite{effblog}. 
A technical report by  Manousis et al.~\cite{manousis2016shedding} analyzed adoption of \LE in May 2016, covering the geo-location of certified domains and use of \LE certificates in malware and typosquatting domains.

To the best of our knowledge, this is the first 
work that singles out \LE and shows what segments of the market are using and deploying their certificates. We demonstrate that \LE is democratizing encryption, by being used mostly by the lower-cost share of the hosting market.

%

\section{Conclusions and Future Work}
\label{sec:conclusions}

\LE has been successful in disrupting the certificate industry, which has been slow in covering the lower-cost end of the market. By addressing the two major barriers inhibiting ubiquitous encryption (cost and complexity required in issuing X.509 certificates), \LE has become one of the largest CAs within only one year after its first certificate was issued.






In this paper, we have studied the certificate issuance in the first year of \LE and showed that it has been playing a major role in democratizing encryption: \LE has been widely used, and mostly by the low-cost share of the market (shared hosting), which would be unlike to deploy the complex and costly X.509 certificates before \LEs.
 
We have also shown that once these barriers are eliminated, it enables big hosting providers to issue and deploy certificates for their customers in bulk, thus quickly and automatically enable encryption across a large number of domains. For example, we have shown that currently, 47\% of \LE certified domains are hosted at three large hosting companies (Automattic/wordpress.com, Shopify, and OVH). 

The success of \LE can also be measured by the fact that 70\% of the \LE certified domains remain active 
after the first issuance of the certificate (\LE certificates expire after three months if not renewed). Also, for one TLD zone (\url{.nl}), we show that \LE certificates have been issued not only for newly registered domains, but also for several-year-old domains, likely benefiting from bulk issuing by their hosting companies.



Issuing a certificate is only one part of the story for encrypted communications: deploying it on the server side is also of essence. To measure the fraction of deployed \LE certificates, we actively scanned a sample of 25K FQDNs. We showed that 63\% of them are correctly deployed for \texttt{https}, which is a lower bound value given that these certificates can also be used for other applications.

As future work, it is important to observe how \LE evolves and how it impacts the CA market, 
and how cyber criminals use malicious domains certified by \LEs.

\vspace{0.1cm}
\textbf{Acknowledgements }{\small
The authors would like to thank J.C. Jones (Mozilla) and 
Carlos Ga\~n{\'a}n (TU Delft) for their valuable comments on paper drafts and Maarten Wullink (SIDN) for sharing his https/DNS scanner (DNS-EMAP).}
\vspace{-0.3cm}

\small

\bibliography{rfc,paper}
\bibliographystyle{abbrv}

\end{document}